\documentclass[12pt]{article}
\usepackage{amssymb,amsmath,epsfig,cite}
\allowdisplaybreaks

\begin{document}
\title{\bf Measure of Complexity in Self-Gravitating Systems using Structure Scalars}
\author{Z. Yousaf$^1$ \thanks{zeeshan.math@pu.edu.pk}, Kazuharu Bamba$^2$
\thanks{bamba@sss.fukushima-u.ac.jp}, M. Z. Bhatti$^3$ \thanks{mzaeem.math@pu.edu.pk} K. Hassan$^4$ \thanks{komalhassan3@gmail.com}\\
$^{1,3,4}$ Department of Mathematics, University of the Punjab,\\
Quaid-i-Azam Campus, Lahore-54590, Pakistan\\
$^2$ Division of Human Support System,\\ Faculty of Symbiotic Systems Science,\\ Fukushima University, Fukushima 960-1296, Japan}

\date{}

\maketitle
\begin{abstract}
The aim of this paper is to present the definition of complexity for
static self-gravitating anisotropic matter proposed in $f(G,T)$
theory, where $G$ is the Gauss-Bonnet term and $T$ is the trace of
energy momentum tensor. We evaluate field equations,
Tolman-Oppenheimer-Volkoff equation, mass functions and structure
scalars. Among the calculated modified scalar variables that are
obtained from the orthogonal splitting of Riemann tensor, a single
scalar function has been identified as the complexity factor. After
exploring the corresponding Tolmann mass function, it is seen that
the complexity factor along with the $f(G,T)$ terms have greatly
influenced its formulation and its role in the subsequent radial
phases of the spherical system. We have also used couple of ansatz
in order to discuss possible solutions of equations of motion in the
study of the structure of compact object.
\end{abstract}

\section*{I. Introduction}

A system is a collection of ordered and interrelated elements. A
minor/major disturbances among the elements of such a system may
cause complexity. Many researchers put forward different definitions
for complexity. Initially, entropy and information were considered a
major criteria to check complexity. To scrutinize the complexity in
mathematical physics, perfect crystal and ideal gas were considered.
Atoms in perfect crystal are completely organized and have symmetry
throughout the structure. A small portion completely describes the
whole notion of the system, in that criterium it provide less
information. While in ideal gas, all particles are randomly
distributed and give maximum information by considering a small
section. This analysis disclosed the fact that order and structure
manifest no complexity. There must be some other factors.

L\'{o}pez-Ruiz \emph{et al.} \cite{lopez1995statistical} put forward
a probabilistic and disequilibrium approach to describe the
complexity of a physical system. It has many applications regarding
different physical systems. Further, for different physical
situations manipulation of complexity becomes easier to work with.
This could not help in case of ideal gas and perfect crystal because
complexity vanishes. Calbet and L\'{o}pez-Ruiz
\cite{calbet2001tendency} formulated the time evolution equations of
``tetrahedral gas". Under some constraints, the gas grows and
maximum complexity occurs in phase space analysis at equilibrium.
Expansion of statistical measure of complexity towards continuous
system was examined by Catalan\emph{et al.}
\cite{catalan2002features} and it needs certain requirements.

Astrophysical objects have components like energy density, pressure,
luminosity which can be utilized to measure complexity. The
$C_{LMC}$ complexity is being calculated by the density of an
astrophysical object, like white dwarf \cite{sanudo2009complexity}
and Chatzisavvas \emph{et. al} \cite{chatzisavvas2009complexity}
then applied this concept to neutron stars. To study evolution of
the dynamical object, Herrera \cite{herrera2018definition}
introduced the  condition of minimal complexity in addition to
complexity factor. Furthermore, the same author
\cite{herrera2019complexity} put forward the concept of complexity
factor to axially symmetric static sources. Herrera \emph{et al.}
\cite{herrera2009structure} obtained that five different scalar
quantities from the orthogonal splitting of Riemann tensor exist in
order to study the structure and evolution of self-gravitating
spherical fluids. Yousaf \emph{et al.}
\cite{doi:10.1139/cjp-2017-0214,yousaf2016influence,Yousaf2019}
also found same number variables in the realm of modified gravity.
They have also checked their role in the modeling Raychaudhari
equations. Herrera \emph{et al.}
\cite{herrera2011physical,PhysRevD.79.087505,herrera2011tilted,herrera2017gibbs}
studied the role of these scalar variables in the emergence of
energy density inhomogeneity on the surface of anisotropic
self-gravitating matter distributions. Bhatti \emph{et al.}
\cite{yousaf2018some,yousaf2016causes}
extended these results and examined that under some constraints the
Weyl scalar is also responsible for producing inhomogeneities over
the regular distribution of fluid configurations.

Herrera \emph{et al.} presented a new definition of complexity
factor the static \cite{herrera2018new} and non-static
\cite{herrera2018definition} spherically symmetric spacetimes.
Sharif and Butt \cite{sharif2018complexity} extended these results
for the static cylindrical spacetimes. Herrera \emph{et al.}
\cite{herrera2019complexity} modified their own results for those
relativistic systems that has axially symmetric geometric
distributions. Yousaf  \cite{yfuture} also calculated such factor by
taking corrections of Palatini $f(R)$ gravity and found that one can
study the complexity of the spherical structure with the help of
structure scalars. recently, Herrera \emph{et al.}
\cite{herrera2019complexityb} performed this analysis by taking
Bondi metric and calculated a particular relation between the
measure of complexity and vorticity of the system. They also
formulated a way to differentiate natural and non-natural
non-dissipative systems. Yousaf \emph{et al.}
\cite{yousaf2020study,yousaf2020measure} modified the definition of
complexity factor in $f(R,T,R_{\mu\nu}T^{\mu\nu})$ gravity and
calculated $Y_{TF}$ as the corresponding complexity factor for the
static relativistic systems.

Anisotropy and inhomogeneous energy density could play an important
role in the study of the complexity factor. Bayin
\cite{bayin1982anisotropic} presented the analytical solutions to
study effects of anisotropy on the structure of compact objects, he
also studied the rotating and radiating anisotropic spheres. When
the system departs from the equilibrium, cracking appeared in
self-gravitating spheres due to the inclusion of anisotropy
\cite{herrera1992cracking}. For self gravitating anisotropic fluid,
Patel and Vaidya \cite{patel1996anisotropic} proposed four exact
analytical solutions. Mak and Harko \cite{mak2002exact} proposed the
exact analytical solution of static anisotropic quark, he deduced
that due to the involvement of anisotropic pressure, mass and radius
of quark star increase. Pinheiro and Chan
\cite{pinheiro2011radiating} presented a new model with
inhomogeneous energy density to a collapsing star and compared the
behavior of physical characteristics, i.e., pressure, luminosity,
energy, adiabatic index with homogeneous energy model.

General relativity (GR) gave the new directions to explore the
universe by opening new ways of research area. No doubt, in the last
century GR has obtained a great achievement in the study of
relativistic structures and universe evolution. Baryon Acoustic
oscillations, clusters of galaxies and Type Ia supernovae are some
of the most important confirmations for the accelerated expansion of
universe
\cite{pietrobon2006integrated,giannantonio2006high,riess2007new}.
Recently, Baryon Oscillation Spectroscopic Survey (BOSS) measured
the power spectrum and angle-averaged galaxy correlation function of
galaxy cluster under the reconstruction of baryon acoustic
oscillation (BAO) feature and found the most accurate distance
constraint which are in agreement with the present supernova
measurements \cite{anderson2012clustering}. From all these
evidences, we have come to the point that there is some kind of
enigmatic energy known as dark energy which could be the reason for
this cosmic expansion. There exist several reviews on the issue of
dark energy \cite{copeland2006dynamics, bamba2012dark} and modified
gravity theories
\cite{nojiri2011unified,nojiri2017modified,capozziello2010beyond,capozziello2011extended,
de2010f,joyce2015beyond,cai2016f,bamba2015inflationary,shamir2019behavior}
as an explanation for the cosmic acceleration.

To study the latest cosmological model and the role of dark
energy/matter in the evolution of our cosmos, researchers used some
other theories which could be called as modified theories. In these
theories, the generic functions of scalar invariants are being added
or replaced in place of Ricci scalar $R$ in the Einstein-Hilbert
action. First modification in GR was done by replacing $R$ with
generic function $f(R)$. Lovelock gravity is generalization of GR in
n-dimensional space \cite{deruelle1990lovelock}. $\Re$ is first
lovelock scalar and Gauss-Bonnet (GB) invariant is the second
lovelock scalar. The term GB consists of Riemann tensor, Ricci
tensor and Ricci scalar embroidered in a special way as $G=R^{\mu\nu
lm}R_{\mu\nu lm}-4R^{\mu\nu}R_{\mu\nu}+R^{2}$.

Nojiri and Odintsov \cite{nojiri2005modified} used $f(G)$ gravity in
order to study late-time cosmological aspects. Late-time era,
coincidence problem and presence of dark matter with inhomogeneous
equation of state are being  discussed with $f(G),~f(R)$ and
$f(R,G)$ models by the same authors \cite{nojiri2007introduction}.
The weak energy condition is established to analyze some pragmatic
$f(G)$ models by utilizing values of hubble, jerk and snap
parameters \cite{garcia2011energy}. To study late-time cosmic
acceleration, Felice and Tsujikawa \cite{de2009solar} considered
$f(G)$ gravity model as one of the best candidate because it
satisfies the solar system constraints. Bamba \emph{et al.}
\cite{bamba2017energy} explored some viability bounds on some
specific $f(G)$ gravity models through numerical technique with the
help of Hubble and snap parameters.

Recently, a new modified theory $f(G,T)$ was introduced by adding
the trace of energy momentum tensor in the action. Sharif and Ikram \cite{sharif2016energya}
used reconstruction process with the help of power-law model in
$f(G,T)$ theory and presented few energy conditions. Some of the
differences in $f(G,T)$ theory in comparison to $GR$ are: the
existence of an extra force, due to which particles pursue
non-geodesic path and force is orthogonal to four velocity of the
fluid. Also, there in non-minimal coupling between matter and $G$
terms, due to which we analyze the non-conserved energy-momentum
tensor. Cosmologically possible $f(G,T)$ gravity models were being
created by Noether symmetry approach \cite{shamir2017noether}. Moreover, for spherically symmetric metrice, gravastars solutions are listed by Shamir and  Ahmad \cite{shamir2018gravastars}. Non-static curvature-matter coupling is
assumed to analyze the shear, Raychaudhuri, and Weyl scalar
equations. Also, Newman-Penrose formalism and Penrose-Hawking
singularity theorems  are studied best by $f(G,T)$ scalar functions
\cite{yousaf2019role}. The same author worked on the scalar
structure in $f(G,T)$ gravity \cite{yousaf2018structure}. Recently, Yousaf et al. \cite{yousaf2020complexity}
calculated the complexity factor for the charged relativistic spherical systems in modified gravity.

In this paper, we study the role of complexity factor in the
modeling of stellar structure in the background of $f(G,T)$ gravity.
The outline of this paper is as follows. In Sec. II, we set up
physical variables and field equations in the presence of $f(G,T)$
gravity. We have also expressed our field equations with the help of
Tolman mass function. Sec. III is devoted to the orthogonal
splitting of Riemann curvature tensor. This gives rise to set of
four structure variables , among them one has been identified as a
complexity factor. In Sec. IV, two proposed solutions of field
equations are examined by vanishing complexity factor. Lastly, we
conclude the results in Sec. V.

\section*{II. The Physical Variables And the Field Equations}

In this section, we will describe the physical variables and its corresponding modified field equations to study the anisotropic self-gravitating fluid.

\subsection*{A. Field Equations}

The action integral to formulate the field equation in $f(G,T)$ gravity is given by
\begin{equation}\label{1a}
 S_{f(G,T)} =\frac{1}{2k^2}\int[f(G,T)+R]\sqrt{-g}d^{4}x+\int\pounds_{m}\sqrt{-g}d^{4}x,
\end{equation}
Here, $\pounds_{m}$ and $g$ serve as the lagrangian density and
determinant of the metric tensor, respectively. The energy momentum
tensor is described as
\begin{equation}\label{75a}
  T_{\mu\nu}= -\frac{2}{\sqrt{-g}}\frac{\delta \sqrt{-g}\pounds_{m}}{\delta g^{\mu\nu}}
\end{equation}
 The following field equations for $f(G,T)$ are obtained by varying Eq.\eqref{1a} with respect to metric tensor as
\begin{eqnarray}\nonumber
G^{\mu}_{\nu}&=&8\pi T^{\mu}_{\nu}-(\Theta^{\mu}_{\nu}+T^{\mu}_{\nu})f_{T}(G,T)+\frac{1}{2}\delta^{\mu}_{\nu}f(G,T)\\\nonumber
&+&(4R_{\nu l}R^{\mu l}+4R^{l m}R^{\mu}_{l \nu m}-2RR^{\mu}_{\nu}-2R^{l m n \mu}R_{\nu l m n})f_{G}(G,T)\\\nonumber
&+&(4R^{\mu}_{\nu}\nabla^{2}+4\delta^{\mu}_{\nu}R^{l m}\nabla_{l}\nabla_{m}+2R\nabla^{\mu}\nabla_{\nu}-2\delta^{\mu}_{\nu}R\nabla^{2}
-4R^{\mu l}\nabla_{l}\nabla_{\nu}\\\label{2a}
&-&4R^{l}_{\nu}\nabla^{\mu}\nabla_{l}-4R^{\mu}_{l \nu m}\nabla^{l}\nabla^{m})f_{G}(G,T),
\end{eqnarray}
where $G_{\mu\nu}=R_{\mu\nu}-\frac{1}{2}Rg_{\mu\nu}$ represents the Einstein tensor and $\nabla^{2}=\Box=\nabla^{l}\nabla_{l}$  describe
the d' Alembert operator and $\Theta_{\mu\nu}=-2T_{\mu\nu}-Pg_{\mu\nu}$. Also, the partial derivatives of $f(G,T)$ w.r.t $G$ and $T$ are represented by $f_{G}(G,T)$ and $f_{T}(G,T)$, respectively. It is worthy to stress that like in $f(R,T)$ theory \cite{harko2011f}, we have used $\pounds_{M}=-P$ in the derivation of our field equations. The detailed derivation of these equations are described by Yousaf \cite{yousaf2019role}. When $f(G,T)=f(G)$, the field equations reduce to $f(G)$ gravity.

One can write Eq.\eqref{2a} in an alternative form as follows
\begin{equation}\label{3a}
  G^{\mu}_{\nu}=8\pi T^{\mu(eff)}_{\nu}=8\pi(T^{\mu(M)}_{\nu}+T^{\mu(GT)}_{\nu}).
\end{equation}
The aim of this paper is to explore the effects of $f(G,T)$ corrections on the definition of complexity factor for the static anisotropic spherically symmetric geometric distribution. Therefore, the source of the gravitation ($T^{\mu(M)}_{\nu}$ mentioned in Eq.\eqref{3a}) is assumed to be anisotropic matter given as follows
\begin{equation}\label{4a}
  T^{\mu(M)}_{\nu}=\mu u^{\mu}u_{\nu}-Ph^{\mu}_{\nu}+\Pi^{\mu}_{\nu},
\end{equation}
where
\begin{equation}\label{5a}
\begin{split}
\Pi^{\mu}_{\nu}=\Pi\left(s^{\mu}s_{\nu}+\frac{1}{3}h^{\mu}_{\nu}\right); \quad  P=\frac{P_{r}+2P_{\bot}}{3}\\*
\Pi=P_{r}-P_{\bot}; \quad  h^{\mu}_{\nu}=\delta^{\mu}_{\nu}-u^{\mu}u_{\nu}.
\end{split}
\end{equation}
The line element for spherically symmetric metric is as follows
\begin{equation}\label{10a}
ds^{2}=e^{\nu}dt^{2}-e^{\lambda}dr^{2}-r^{2}[d\theta^{2}+{sin^{2}\theta}{d^{2}\phi}],
\end{equation}
and $s^{\mu}$ has the representation
\begin{equation}\label{6a}
s^{\mu}=\left(0,e^{\frac{-\lambda}{2}},0,0\right),
\end{equation}
satisfying the properties $s^{\mu}u_{\mu}=0$, $s^{\mu}s_{\mu}=-1$.
In the energy-momentum tensor, four velocity vector $u^{\mu}$ is written as
\begin{equation}\label{7a}
u^{\mu}=\left(e^{-\frac{\nu}{2}},0,0,0\right).
\end{equation}
We can calculate the four acceleration from it as
$a^{\alpha}=u^{\alpha}_{;\beta}u^{\beta}$
and it is found that it has only one non-vanishing component which is
\begin{equation}\label{8a}
a_{1}=-\frac{\nu'}{2},
\end{equation}
and the modified terms of $f(G,T)$ are defined as
\begin{eqnarray}\nonumber
T^{\mu(GT)}_{\nu}&=&\frac{1}{8\pi}\left[\{(\mu+P)u^{\mu}u_{\nu}+\Pi^{\mu}_{\nu}\}f_{T}(G,T)+\frac{1}{2}\delta^{\mu}_{\nu}f(G,T)\right.\\\nonumber
&+&\left.(4R_{\nu l}R^{\mu l}+4R^{l m}R^{\mu}_{l \nu m}-2RR^{\mu}_{\nu}-2R^{l m n \mu}R_{\nu l m n})f_{G}(G,T)\right.\\\nonumber
&+&\left.(4R^{\mu}_{\nu}\nabla^{2}+4\delta^{\mu}_{\nu}R^{l m}\nabla_{l}\nabla_{m}+2R\nabla^{\mu}\nabla_{\nu}-2\delta^{\mu}_{\nu}R\nabla^{2}
-4R^{\mu l}\nabla_{l}\nabla_{\nu}\right.\\\label{9a}
&-&\left.4R^{l}_{\nu}\nabla^{\mu}\nabla_{l}-4R^{\mu}_{l \nu m}\nabla^{l}\nabla^{m})f_{G}(G,T)\right].
\end{eqnarray}

The corresponding $f(G,T)$ field equations are
\begin{eqnarray}\label{11a}
  \mu^{eff} &=& -\frac{1}{8\pi}\left[-\frac{1}{r^{2}}+e^{-\lambda}(\frac{1}{r^{2}}-\frac{\lambda'}{r})\right], \\\label{49a}
  P_{r}^{eff} &=&  -\frac{1}{8\pi}\left[-\frac{1}{r^{2}}+e^{-\lambda}(\frac{1}{r^{2}}+\frac{\nu'}{r})\right],  \\\label{50a}
 P_{\bot}^{eff}&=&\frac{1}{32\pi}\left(2\nu''+\nu'^{2}-{\lambda'}{\nu'}+2\frac{\nu'-\lambda'}{r}\right),
\end{eqnarray}
where prime denotes the derivatives with respect to $r$. The values of $ \mu^{eff}$, $P_{r}^{eff}$ and $P_{\bot}^{eff}$ are given in the Appendix of \cite{yousaf2020complexity}.

The covariant divergence of Eq.\eqref{2a} is non-zero and is found to be
\begin{eqnarray}\nonumber
\nabla^{\mu}T_{\mu\nu}=\frac{f_{T}(G,T)}{k^{2}-f_{T}(G,T)}\left[(T_{\mu\nu}+\Theta_{\mu\nu})\nabla^{\mu}(lnf_{T}(G,T))
\right.\\\label{co1}\left.\nabla^{\mu}\Theta_{\mu\nu}-\frac{1}{2}g_{\mu\nu}\nabla^{\mu}T\right].
\end{eqnarray}
The non-conserved hydrostatic equilibrium equation in $f(G,T)$ gravity can be derived by using the Eqs.\eqref{11a}-\eqref{50a} as
\begin{equation}\label{12a}
 P^{'eff}_{r}=-\frac{\nu'}{2}(P^{eff}_{r}+\mu^{eff})+2\frac{(P^{eff}_{\bot}-P^{eff}_{r})}{r}+e^{\lambda} Z,
 \end{equation}
where value of $Z$ is given in Appendix of \cite{yousaf2020complexity}. Equation \eqref{12a} is known as Tolman-Oppenheimer-Volkoff equation for anisotropic fluid.

From Eq.\eqref{49a}, one can get the value of the metric coefficient as follows
\begin{equation}\label{13a}
 \nu'=2\frac{m+4\pi r^{3}P^{eff}_{r}}{r(r-2m)},
\end{equation}
where $m$ is the mass function that can be expressed as
\begin{equation}\label{15a}
  1-e^{-\lambda}=\frac{2m}{r},
\end{equation}
One can write Eq.\eqref{12a} after using Eq.\eqref{13a} as follows
\begin{equation}\label{14a}
P^{'eff}_{r}=-\frac{m+4\pi r^{3}P^{eff}_{r}}{r(r-2m)}(\mu^{eff}+P^{eff}_{r})+2\frac{(P^{eff}_{\bot}-P^{eff}_{r})}{r}+e^{\lambda} Z,
\end{equation}
With the help of Eqs.\eqref{11a}-\eqref{50a}, Eq. \eqref{15a} can also be written as
\begin{equation}\label{16a}
m=4\pi\int_{0}^{r}\tilde{r}^{2}\mu^{eff}d\tilde{r}.
\end{equation}
Now we assume that a 3-dimensional hypersurface $\Sigma$ has differentiated our manifolds into two different portions. The one is known as an interior one whose manifold has been described already with the help of static spherically symmetric spacetime given in Eq.\eqref{10a}. The other portion also known as the exterior geometry, can be defined with the help of the following Schwarzschild metric as
\begin{equation}\label{17a}
ds^{2}=\left(1-\frac{2M}{r}\right)dt^{2}-\frac{dr^{2}}{\left(1-\frac{2M}{r}\right)}-r^{2}\left(d\theta^{2}+sin^{2}\theta d\phi^{2}\right).
\end{equation}
Following Darmois \cite{darmois1927memorial}, we need first and second conditions of continuity to match the two metrics \eqref{10a} and \eqref{17a} smoothly and gently at boundary surface $\Sigma$. At the boundary surface, these give the following constraints
\begin{eqnarray}\label{18a}
  e^{\nu_{\Sigma}} &=& \left(1-\frac{2M}{r_{\Sigma}}\right), \\\label{65a}
  e^{-\lambda_{\Sigma}} &=&\left(1-\frac{2M}{r}\right),  \\\label{66a}
  \left[ P^{eff}_{r}\right]_{\Sigma}&=&\chi_{1},
 \end{eqnarray}
where subscript $\Sigma$ demonstrates that the subsequent values are calculated at $\Sigma$, while the expression for $\chi_{1}$ is given in Appendix of \cite{yousaf2020complexity}. It is worthy to stress that we have taken $f_{G}=\tilde{f}_{G}$,$f_{T}=\tilde{f}_{T}$,$f=\tilde{f}$ while calculating the above mentioned constraints at the hypersurface. Such constraints need to be valid at the boundary for the smooth matching of manifolds. Such kind of condition with their detailed proof have been describe by Yousaf et al. \cite{Yousaf2017,yousaf2017stability} for the $f(R,T,R_{\mu\nu}T^{\mu\nu})$ theory of gravity. The manifolds \eqref{10a} and \eqref{17a} are smoothly matched at boundary if and only if above three conditions \eqref{18a}, \eqref{65a} and \eqref{66a} are satisfied.

\subsection*{B. The Riemann Curvature and the Weyl tensor}

The Weyl tensor has zero magnetic part for the spherical symmetric case. So it is expressed only in electric part $\left(E_{\alpha\beta}=C_{\alpha\gamma\beta\delta}u^{\gamma}u^{\delta}\right)$ with
\begin{equation}\label{20a}
C_{\mu\nu\kappa\lambda}=\left(g_{\mu\nu\alpha\beta}g_{\kappa\lambda\gamma\delta}-\eta_{\mu\nu\alpha\beta}\eta_{\kappa\lambda\gamma\delta}\right)
u^{\alpha}u^{\gamma}E^{\beta\delta},
\end{equation}
where $g_{\mu\nu\alpha\beta}=g_{\mu\alpha}g_{\mu\beta}-g_{\mu\beta}g_{\nu\alpha}$ ,and $\eta_{\mu\nu\alpha\beta}$ denotes the Levi-Civita tensor. $E_{\alpha\beta}$ can also be written as
\begin{equation}\label{21a}
E_{\alpha\beta}=E\left(s_{\alpha}s_{\beta}+\frac{1}{3}h_{\alpha\beta}\right),
\end{equation}
with
\begin{equation}\label{22a}
E=-\frac{e^{-\lambda}}{4}\left[\nu''+\frac{\nu'^{2}-\lambda'\nu'}{2}-\frac{\nu'-\lambda'}{r}+\frac{2\left(1-e^{\lambda}\right)}{r^{2}}\right],
\end{equation}
has the properties $E^{\alpha}_{\alpha}=0,~E_{\alpha\gamma}=E_{(\alpha\gamma)}$ and $E_{\alpha\gamma}u^{\gamma}=0$.

\subsection*{C. The Mass Function And the Tolman Mass}

Here we are going to discuss the mass of an interior sphere at the surface $\Sigma$ in two different forms and their relation with the Weyl tensor.

\subsubsection*{The Mass Function}

From Eqs.\eqref{3a}, \eqref{15a} and \eqref{21a}, we can write the mass function as
\begin{equation}\label{23a}
m=\frac{4\pi}{3}r^{3}\left(\mu^{eff}+P^{eff}_{\bot}-P^{eff}_{r}\right)+\frac{r^{3}E}{3},
\end{equation}
from which $E$ can be evaluated as
\begin{equation}\label{24a}
E=-\frac{4\pi}{r^{3}}\int_{0}^{r}\tilde{r}^{3}\mu'^{eff}d\tilde{r}+4\pi\left(P^{eff}_{r}-P^{eff}_{\bot}\right).
\end{equation}
After inserting Eq.\eqref{24a} in Eq.\eqref{23a}, the mass function can be rewritten as
\begin{equation}\label{25a}
m(r)=\frac{4\pi}{3}r^{3}\mu^{^{eff}}-\frac{4\pi}{3}\int_{0}^{r}\tilde{r}^{3}\mu'^{eff}d\tilde{r}.
\end{equation}
 Equation \eqref{24a} shows the relation of the Weyl tensor with two physical properties of the fluid, i.e. density inhomogeneity and local anisotropic pressure, whereas in Eq. \eqref{25a} mass function has been expressed in terms of homogeneous energy density and radial variations caused by density inhomogeneity in the presence of $f(G,T)$ dark source terms.

\subsubsection*{Tolman Mass in $f(G,T)$ gravity}

The study to understand the effects of matter variables on the formulation of Tolman mass has gained utmost importance. Herrera \emph{et al.} \cite{herrera1997role,herrera2011role} expressed the Tolman mass with the help of fluid matter variables. They also described the role of this mass function on the pace of spherical collapse. Bhatti \emph{et al.} \cite{zaeem2019energy} modified their results and defined the same function in $f(R)$ and Maxwell-$f(R)$ theory of gravity for the relativistic spherical structures Here, we can write it as \cite{tolman1930use}
\begin{equation}\label{26a}
m_{T}=4\pi\int^{r_{\Sigma}}_{0}r^{2}e^{\frac{\nu+\lambda}{2}}\left(T^{0(eff)}_{0}-T^{1(eff)}_{1}-2T^{2(eff)}_{2}\right)dr.
\end{equation}
As total energy of the fluid is measured in the form of Tolman formula, so the mass inside boundary $\Sigma$ of radius $r$ can be followed as
\begin{equation}\label{27a}
m_{T}=4\pi\int^{r}_{0}\tilde{r}^{2}\left(T^{0(eff)}_{0}-T^{1(eff)}_{1}-2T^{2(eff)}_{2}\right)d\tilde{r}.
\end{equation}
The ``active gravitational mass" plays an important role from the global concept of energy to the local level, which can seen from below as
\begin{equation}\label{28a}
m_{T}=\left[m(r)+4\pi r^{3}P^{eff}_{r}\right].
\end{equation}
Alternatively, by using field equations \eqref{11a}, \eqref{49a} and \eqref{50a}, we have
\begin{equation}\label{29a}
m_{T}=e^{\frac{\nu-\lambda}{2}}\nu'\frac{r^{2}}{2}.
\end{equation}
The gravitational acceleration $(a=-s^{\nu}a_{\nu})$ of a test particle is obtained through Eq.\eqref{18a} as
\begin{equation}\label{30a}
a=\frac{e^{\frac{-\lambda}{2}}\nu'}{2}=\frac{e^{\frac{-\nu}{2}}m_{T}}{r^{2}}.
\end{equation}
Another way to write $m_{T}$ is
\begin{eqnarray}\nonumber
m_{T}&=&(m_{T})_{\Sigma}\left(\frac{r}{r_{\Sigma}}\right)^{3}\\\label{31a}&-&r^{3}\int^{r_{\Sigma}}_{r}e^{\frac{\nu+\lambda}{2}}\left[\frac{8\pi}
{\tilde{r}}\left(P^{eff}_{\bot}-P^{eff}_{r}\right)+\frac{1}{\tilde{r}^{4}}\int^{\tilde{r}}_{0}4\pi \tilde{r}^{3}\mu'^{eff}d\tilde{r}\right]d\tilde{r}.
\end{eqnarray}
By making use of Eq.\eqref{24a}, we may write above equation as
\begin{equation}\label{32a}
m_{T}=(m_{T})_{\Sigma}\left(\frac{r}{r_{\Sigma}}\right)^{3}-r^{3}\int^{r_{\Sigma}}_{r}\frac {e^{\frac{\nu+\lambda}{2}}}{\tilde{r}}\left[4\pi\left(P^{eff}_{\bot}-P^{eff}_{r}\right)-E\right]d\tilde{r}.
\end{equation}
In Eq. \eqref{31a}, the $2nd$ integral express $m_{T}$ in the form of anisotropic pressure and inhomogeneous energy density in the presence of $f(G,T)$ dark source terms. Now we move forward to the orthogonal splitting of Riemann tensor.

\section*{III. The Orthogonal Splitting of the Riemann Tensor in $f(G,T)$ Garvity}

For the first time Bel \cite{bel1961inductions} introduced the orthogonal splitting of Riemann tensor. Herrera et al. \cite{herrera2004spherically,herrera2011role} described the orthogonal splitting of Riemann tensor and elaborated this technique in the modeling
of radiating and non-radiating stellar structure in GR. Now, after some alteration, one can write the following tensors as
\begin{eqnarray}\label{33a}
  Y_{\alpha\beta} &=&R_{\alpha\gamma\beta\delta}u^{\gamma}u^{\delta},  \\
  Z_{\alpha\beta} &=&_{\ast}R_{\alpha\gamma\beta\delta}u^{\gamma}u^{\delta}=\frac{1}{2}\eta_{\alpha\gamma\epsilon\mu}R^{\epsilon\mu}_{\beta\delta}u^{\gamma}u^{\delta},  \\
  X_{\alpha\beta} &=&_{\ast}R^{\ast}_{\alpha\gamma\beta\delta}u^{\gamma}u^{\delta}=\frac{1}{2}\eta^{\epsilon\mu}_{\alpha\gamma},
  R^{\ast}_{\epsilon\mu\beta\delta}u^{\gamma}u^{\delta},
\end{eqnarray}
Here the dual tensor is represented by $\ast$, i.e., $R^{\ast}_{\alpha\beta\gamma\delta}=\frac{1}{2}\eta_{\epsilon\mu\gamma\delta}R^{\epsilon\mu}_{\alpha\beta}$.
Now we orthogonally split the Riemann tensor. For that purpose, we can write the formula of Riemann tensor with the help of field equations as
\begin{equation}\label{34a}
R^{\alpha\gamma}_{\beta\delta}=C^{\alpha\gamma}_{\beta\delta}+16\pi T^{eff[\alpha}_{[\beta}\delta^{\gamma]}_{\delta]}+8\pi
 T^{(eff)}\left(\frac{1}{3}\delta^{\alpha}_{[\beta}\delta^{\gamma}_{\delta]}-\delta^{[\alpha}_{[\beta}\delta^{\gamma]}_{\delta]}\right),
\end{equation}
When we insert Eqs.\eqref{4a} and \eqref{9a} in Eq.\eqref{34a}, one can get the following tensorial quantites
\begin{eqnarray}\nonumber
 R^{\alpha\gamma}_{(I)\beta\delta} &=& 16\pi\mu\mu^{[\alpha}\mu_{[\beta}\delta^{\gamma]}_{\delta]}+2\mu\mu^{[\alpha}\mu_{[\beta}\delta^{\gamma]}_{\delta]}
 -16\pi Ph^{[\alpha}_{[\beta}\delta^{\gamma]}_{\delta]}\\\label{70a}&+&2P \mu^{[\alpha}
 \mu_{[\beta}\delta^{\gamma]}_{\delta]}+8\pi(\mu-3P)\left(\frac{1}{3}\delta^{\alpha}_{[\beta}\delta^{\gamma}_{\delta]}-
 \delta^{[\alpha}_{[\beta}\delta^{\gamma]}_{\delta]}\right) \\\nonumber
 R^{\alpha\gamma}_{(II)\beta\delta} &=& 16\pi\Pi^{[\alpha}_{[\beta}\delta^{\gamma]}_{\delta]}+2\Pi^{[\alpha}_{[\beta}\delta^{\gamma]}_{\delta]}+
 \delta^{[\alpha}_{[\beta}\delta^{\gamma]}_{\delta]}f\\\label{71a}&+&8\delta^{[\alpha}_{[\beta}\delta^{\gamma]}_{\delta]}
 R^{lm}\nabla_{l}\nabla_{m}f_{G}-4R\delta^{[\alpha}_{[\beta}\delta^{\gamma]}_{\delta]}\Box f_{G}  \\\label{72a}
 R^{\alpha\gamma}_{(III)\beta\delta} &=& 4\mu\mu^{[\alpha}_{[\beta}E^{\gamma]}_{\delta]}-\epsilon^{\alpha\gamma}_{\mu}\epsilon_{\beta\delta\nu}E^{\mu\nu} \\\nonumber
 R^{\alpha\gamma}_{(IV)\beta\delta} &=&2(R_{l\beta}R^{l\alpha}\delta^{\gamma}_{\delta}-R_{l\delta}R^{l\alpha}\delta^{\gamma}_{\beta}+R_{l\beta}R^{l\gamma}
 \delta^{\alpha}_{\delta}+R_{l\delta}R^{l\gamma}\delta^{\alpha}_{\beta})f_{G}\\\nonumber&+&2R^{lm}(R^{\alpha}_{l\beta m}\delta^{\gamma}_{\delta}
-R^{\alpha}_{l\delta m}\delta^{\gamma}_{\beta}-R^{\gamma}_{l\beta m}\delta^{\alpha}_{\delta}+R^{\gamma}_{l\delta m}\delta^{\alpha}_{\beta})f_{G}\\\nonumber&-&
R(R^{\alpha}_{\beta}\delta^{\gamma}_{\delta}-R^{\alpha}_{\delta}\delta^{\gamma}_{\beta}-R^{\gamma}_{\beta}\delta^{\alpha}_{\gamma}
+R^{\gamma}_{\delta}\delta^{\alpha}_{\beta})f_{G}\\\nonumber&-&2(R^{\alpha}_{l\beta m}\delta^{\gamma}_{\delta}-R^{\alpha}_{l\delta m}\delta^{\gamma}_{\beta}-R^{\gamma}_{l\beta m}\delta^{\alpha}_{\delta}
+R^{\gamma}_{l\delta m}\delta^{\alpha}_{\beta})\nabla^{l}\nabla^{m}f_{G}\\\nonumber&+&2(R^{\alpha}_{\beta}\delta^{\gamma}_{\delta}-R^{\alpha}_{\delta}\delta^{\gamma}_{\beta}-R^{\gamma}_{\beta}\delta^{\alpha}_{\gamma}
+R^{\gamma}_{\delta}\delta^{\alpha}_{\beta})\Box f_{G}\\\nonumber&+&R(\delta^{\gamma}_{\delta}\nabla^{\alpha}_{\beta}-\delta^{\gamma}_{\beta}\nabla^{\alpha}_{\delta}
-\delta^{\alpha}_{\delta}\nabla^{\gamma}_{\beta}+\delta^{\alpha}_{\beta}\nabla^{\gamma}_{\delta})f_{G}\\\nonumber&-&2(R^{l\alpha}\delta^{\gamma}_{\delta}
\nabla_{\beta}\nabla_{l}-R^{l\alpha}\delta^{\gamma}_{\beta}\nabla_{\delta}\nabla_{l}-R^{l\gamma}\delta^{\alpha}_{\delta}\nabla_{\beta}\nabla_{l}
+R^{l\gamma}\delta^{\alpha}_{\beta}\nabla_{\delta}\nabla_{l})f_{G}\\\nonumber&-&2(R^{l}_{\beta}\delta^{\gamma}_{\delta}\nabla^{\alpha}\nabla_{l}
-R^{l}_{\delta}\delta^{\gamma}_{\beta}\nabla^{\alpha}\nabla_{l}
-R^{l}_{\beta}\delta^{\alpha}_{\delta}\nabla^{\gamma}\nabla_{l}+R^{l}_{\delta}\delta^{\alpha}_{\beta}\nabla^{\gamma}\nabla_{l})f_{G}
\\\nonumber&-&(R_{\beta lmn}R^{lmn\alpha}\delta^{\gamma}_{\delta}-R_{\delta lmn}R^{lmn\alpha}
\delta^{\gamma}_{\beta}-R_{\beta lmn}R^{lmn \gamma}\delta^{\alpha}_{\delta}+R_{\delta lmn}R^{lmn \gamma}\delta^{\alpha}_{\beta})f_{G}
\\\nonumber&+&\frac{1}{3}\left[(\mu+P)f_{T}+2f+4R_{l\nu}R^{l\nu}f_{G}+4R^{lm}R^{\nu}_{l\nu m}f_{G}-2R^{2}f_{G}
  -2R^{\mu}_{lmn}R^{lmn}_{\mu}f_{G}\right.\\\label{73a}&-&\left.2R\Box f_{G}+16R^{lm}\nabla_{l}\nabla_{m}f_{G}
  -4R^{l\nu}\nabla_{\nu}\nabla_{l}f_{G} -4R^{l\mu}\nabla_{l}\nabla_{\mu}f_{G}-4R^{\nu}_{l\nu m}\nabla^{l}\nabla^{m}f_{G}\right],
\end{eqnarray}
with the properties,
\begin{equation}\label{35a}
\epsilon_{\alpha\gamma\beta}=u^{\mu}\eta_{\mu\alpha\gamma\beta},   \quad        \epsilon_{\alpha\gamma\beta}u^{\beta}=0,
\end{equation}
and
\begin{equation}\label{74a}
  \epsilon^{\mu\gamma\nu}\epsilon_{\nu\alpha\beta}=\delta^{\gamma}_{\alpha}h^{\mu}_{\beta}-\delta^{\mu}_{\alpha}h^{\gamma}_{\beta}
  +u_{\alpha}(u^{\mu}\delta^{\gamma}_{\beta}-\delta^{\mu}_{\beta}u^{\gamma}).
\end{equation}
Now we evaluate three tensors $Y_{\alpha\beta},~Z_{\alpha\beta}$ and $X_{\alpha\beta}$ explicitly in form of physical variables as
\begin{eqnarray}\nonumber
  Y_{\alpha\beta} &=& \frac{4\pi}{3} (\mu+3P)h_{\alpha\beta}+\frac{1}{6}(\mu+P)h_{\alpha\beta}f_{T}+E_{\alpha\beta}\\\label{36a}&+&4\pi \Pi_{\alpha\beta}
  +\frac{\Pi_{\alpha\beta}}{2}f_{T}+D^{(D)}_{\alpha\beta},   \\\label{37a}
   X_{\alpha\beta}&=& \frac{8\pi }{3}\mu h_{\alpha\beta}-E_{\alpha\beta}+4\pi \Pi_{\alpha\beta}+\frac{\Pi_{\alpha\beta}}{2}f_{T}+O^{(D)}_{\alpha\beta},
\end{eqnarray}
and
\begin{equation}\label{38a}
Z_{\alpha\beta}=I^{(D)}_{\alpha\beta}.
\end{equation}
The values of dark source terms $D^{(D)}_{\alpha\beta}$, $O^{(D)}_{\alpha\beta}$ and $I^{(D)}_{\alpha\beta}$ are expressed in Appendix of \cite{yousaf2020complexity}.

The above tensors can also be written in form of some scalar functions, referred to as structure scalars. We can define four scalars functions from the tensors $X_{\alpha\beta}$ and $Y_{\alpha\beta}$, these are denoted by $X_{T},~X_{TF},~Y_{T}$,
$Y_{TF}$. A fifth scalar is related to the tensor $Z_{\alpha\beta}$ which disappear here because of the the static spherical case. Energy density inhomogeneity, anisotropic pressure and Tolman mass are some of the elementary characteristics which are described through these scalars \cite{sharif2015structure}. These scalars are found for $f(G,T)$ theory as follows
\begin{eqnarray}\label{39a}
   X_{T}&=& 8\pi \mu+Q^{(D)},  \\\label{40ab}
  X_{TF} &=& 4\pi \Pi-E+\frac{\Pi}{2}f_{T},
\end{eqnarray}
where the expression for $Q^{(D)}$ is given in Appendix of \cite{yousaf2020complexity}. By using Eq.\eqref{24a}
\begin{eqnarray}\label{40a}
  X_{TF} &=& \frac{4\pi}{r^{3}}\int_{0}^{r}\tilde{r}^{3}\mu'^{eff}d\tilde{r}+\frac{\Pi}{2}f_{T}-4\pi\Pi ^{(D)}, \\\label{41a}
  Y_{T} &=& 4\pi\left(\mu+3P_{r}-2\Pi \right)+\frac{1}{2}\left(\mu+P\right)f_{T}+M^{(D)},  \\\label{42a}
  Y_{TF} &=& 4\pi \Pi+E+\frac{\Pi}{2}f_{T}+L^{(D)},
\end{eqnarray}
where $L^{(D)}=\frac{J^{(D)}_{\alpha\beta}}{s_{\alpha}s_{\beta}+\frac{1}{3}h_{\alpha\beta}}$. The values of $M^{(D)}$ and $L^{(D)}$ are described in Appendix of \cite{yousaf2020complexity}. By making use of $E$ from Eq.\eqref{24a}, we obtain
\begin{equation}\label{43a}
Y_{TF}=8\pi \Pi-\frac{4\pi}{r^{3}}\int_{0}^{r}\tilde{r}^{3}\mu'^{eff}d\tilde{r}+\frac{\Pi}{2}f_{T}+4\pi\Pi^{(D)}+L^{(D)}.
\end{equation}
After considering Eqs.\eqref{40a} and \eqref{43a}, we deduce that $X_{TF}$ is responsible for controlling the inhomogeneous energy in the presence of $f(G,T)$ corrections whereas $Y_{TF}$ is involved in the contribution of the anisotropic pressure, energy density inhomogeneity with extra curvature terms. This indicates that $Y_{TF}$ is serving as a complexity factor for our relativistic diagonally symmetric spherical geometry. From these equations, we also deduce that $ X_{TF}$ and $Y_{TF}$ determine the local anisotropy of pressure with some dark source terms.
\begin{equation}\label{44a}
8\pi \Pi+\Pi f_{T}+L^{(D)}= X_{TF}+Y_{TF}.
\end{equation}
To understand $Y_{T}$ and $Y_{TF}$ physically, we go back to Eqs.\eqref{31a} and \eqref{32a}. From Eqs.\eqref{42a} and \eqref{43a}, we obtain
\begin{equation}\label{45a}
m_{T}=(m_{T})_{\Sigma}\left(\frac{r}{r_{\Sigma}}\right)^{3}+r^{3}\int^{r_{\Sigma}}_{r}\frac {e^{\frac{\nu+\lambda}{2}}}{\tilde{r}}
[ Y_{TF}-\frac{\Pi}{2}f_{T}-L^{(D)}+4\pi \Pi^{(D)}]d\tilde{r},
\end{equation}
By comparing the above equation with Eq.\eqref{31a}, it can be seen that Tolman mass in term of anisotropic pressure and inhomogeneous energy density along with dark source terms in $f(G,T)$ gravity is represented by $Y_{TF}$. Alternatively, $Y_{TF}$ explains the modification of Tolman mass as compared to homogeneous matter. Lastly, Tolman mass can also be deduced in the form of
\begin{equation}\label{46a}
m_{T}=\int^{r}_{0}\tilde{r}^{2}e^{\frac{\nu+\lambda}{2}}\left(Y_{T}-\frac{1}{2}(\mu+P)f_{T}+4\pi (\mu^{(D)}+3P^{(D)})-M^{(D)}\right)d\tilde{r}.
\end{equation}
It can be seen from the literature \cite{herrera2004spherically,herrera2011role,yousaf2019role} that the structure scalar, $Y_{T}$ describes the effective form of mass density. In the above expression, it is noticed that $m_{T}$ is expressed with the help of $Y_T$. Thus, $m_T$ could be considered as a possible mathematical vehicle to understand the information as well as the subsequent changes in the matter terms along with the correction terms of $f(G,T)$ gravity.

\section*{IV. Fluid Distributions With Zero Complexity Factor}

The modified equations of motion make a system of three partial differential equations with five unknown functions $(\mu, \nu, \lambda, P^{eff}_{r}, P^{eff}_{\bot})$. In this direction, the vanishing complexity factor condition after applying $Y_{TF}=0$ gives
\begin{equation}\label{47a}
\Pi=\frac{1}{8\pi+\frac{f_{T}}{2}}\left[\frac{4\pi}{r^{3}}\int^{r}_{0}\tilde{r}^{3}\mu'^{eff}d\tilde{r}-L^{(D)}-4\pi \Pi^{(D)}\right].
\end{equation}
It can be seen that the dark source terms reduces the complexity in the vanishing complexity factor condition. After applying the condition $Y_{TF}=0$ one can solve these equations. However, we further need one more condition to solve this system. For this purpose, we proceed our results as follows.

\subsection*{A. The Gokhroo and Mehra Ans\"{a}tz}

Elizalde {et al.} \cite{elizalde2011nonsingular} claimed
that such extra degrees of freedom mediated by exponential $f(R)$
terms may provide a unified picture of our accelerating universe at
both late and early time epochs. To study the behavior of compact
objects, Gokhroo and Mehra \cite{gokhroo1994anisotropic} put forward
solution of the field equations for anisotropic sphere with variable
energy density. As we have merged all the modified geometric
quantities in $\mu^{eff}$, so there would be no change in the
formula arrangement except that now we are dealing the modified
matter. The starting point is a supposition in the form of metric
function $\lambda$ which is
\begin{equation}\label{48a}
e^{-\lambda}=1-\alpha r^{2}+\frac{3K\alpha r^{4}}{5r^{2}_{\Sigma}},
\end{equation}
where $K$ is constant within the range of $(0,1)$ and $\alpha=\frac{8\pi \mu_{o}}{3}$. Equations \eqref{11a} and \eqref{15a} give
\begin{equation}
\mu^{eff}=\mu_{o}\left(1-\frac{Kr^{2}}{r^{2}_{\Sigma}}\right),
\end{equation}
and
\begin{equation}\label{51a}
m(r)=\frac{4\pi \mu_{o}r^{3}}{3}\left(1-\frac{3Kr^{2}}{5r^{2}_{\Sigma}}\right).
\end{equation}
Further, we can write from Eqs.\eqref{49a} and \eqref{50a} as
\begin{equation}\label{52a}
8\pi(P^{eff}_{r}-P^{eff}_{\bot})=e^{-\lambda}\left[\frac{1}{r^{2}}+\frac{\nu'}{2r}-\frac{\nu''}{2}-\left(\frac{\nu'}{2}\right)^{2}
+\frac{\lambda'}{2}\left(\frac{\nu'}{2}+\frac{1}{r}\right)\right]-\frac{1}{r^{2}}.
\end{equation}
To rewrite the line element, we introduce the new variables
\begin{equation}\label{53a}
e^{\nu(r)}=e^{\int(2z(r)-2/r)dr},
\end{equation}
and
\begin{equation}\label{54a}
e^{-\lambda}=y(r),
\end{equation}
and by inserting the Eqs. \eqref{53a} and \eqref{54a} in Eq \eqref{52a} , we obtain
\begin{equation}\label{55a}
y'+y\left[2z+\frac{2z'}{z}+\frac{4}{zr^{2}}-\frac{6}{r}\right]=-\frac{2}{z}\left(\frac{1}{r^{2}}+8\pi \Pi \right),
\end{equation}
Then, the interior spherically symmetric line element becomes
\begin{eqnarray}\nonumber
ds^{2}&=&-e^{\int(2z(r)-2/r)}dt^{2}+\frac{z^{2}e^{\int(2z(r)+\frac{4}{r^{2}z(r)})}}{r^{6}\left(-2\int\frac{z(r)(1+8\pi \Pi r^{2})e^{\int(2z(r)+\frac{4}
{r^{2}z(r)})dr}}{r^{8}}dr+C\right)}dr^{2}\\\label{56}&+&r^{2}d\theta^{2}+r^{2}sin^{2}\theta d\phi^{2},
 \end{eqnarray}
where $C$ is constant of integration. Now, the values of matter variables can written with the help of the above mathematical relations as
 \begin{align}\label{57a}
 4\pi P^{eff}_{r}&=\frac{z(r-2m)+\frac{m}{r}-1}{r^{2}},\\\label{58a}
 4\pi \mu^{eff}&=\frac{m'}{r^{2}},\\\label{59a}
 8\pi P^{eff}_{\bot}&=\left(1-\frac{2m}{r}\right)\left(z'+z^{2}-\frac{z}{r}+\frac{1}{r^{2}}\right)+z\left(\frac{m}{r^{2}}-\frac{m'}{r}\right).
 \end{align}
It can be noticed that the above mentioned results are regular at
the origin $r=0$ obeying $\mu^{(eff)}>0, \mu^{(eff)}>P^{(eff)}_{r},
P^{(eff)}_{\bot}$ relations. The solution must fulfil the junction
conditions to avert the singularity behavior of matter variables.

 \subsection*{B. The Polytrope with Disappearing Complexity Factor}

Self-gravitating fluid consists of different physical variables such
as pressure, density, temperature etc. All these factors have
different effects on the internal structure. The polytropic equation
helps to study the compact objects in a better way. Here we would
use the relation of effective energy density and pressure as we are
going to deal the TOV equation, mass function, fluid with zero
complexity which we have been derived in terms of effective matter
in the scenario of $f(G,T)$ theory. The polytropic equation for
anisotropic matter along with the vanishing complexity condition can
be described as
 \begin{equation}\label{60a}
 P^{eff}_{r}=K\mu^{eff(\gamma)}=K\mu^{eff(1+1/n)};  \quad
        Y_{TF}=0,
   \end{equation}
where constant $K,~\gamma$ and $n$ are generally called as
polytropic constant, polytropic exponent and polytropic index,
respectively. We get two equations from polytropic equation of
state, one of which is obtained from Tolman-Oppenheimer-Volkoff
equation
 \begin{eqnarray}\nonumber
    &&\xi^{2}\frac{d\Psi}{d\xi}\left[\frac{1-2\upsilon \alpha (n+1)/\xi}{1+\alpha\Psi}\right]+\upsilon+\alpha\xi^{3}\Psi^{n+1}+\frac{2\Pi\xi\Psi^{-n}}
    {P^{eff}_{rc}(n+1)}\\\label{61a}&&\times\left[\frac{1-2\upsilon \alpha (n+1)/\xi}{1+\alpha\Psi}\right]=e^{\lambda}Z\left[\frac{\xi^{2}
    (1-2\upsilon \alpha (n+1)/\xi)}{P^{eff}_{rc}A\Psi^{n}(1+\alpha\Psi)(1+n)}\right],
 \end{eqnarray}
 and other is
 \begin{equation}\label{62a}
 \frac{d\upsilon}{d\xi}=\xi^{2}\Psi^{n},
 \end{equation}
It could be beneficial to define few dimensionless variables as follows
 \begin{equation}\label{63a}
 \begin{split}
 \alpha=\frac{P^{eff}_{rc}}{\mu^{eff}_{c}}, \quad  r=\frac{\xi}{A}, \quad   A^{2}=\frac{4\pi\mu^{eff}_{c}}{\alpha(1+n)},\\*
 \Psi^{n}=\frac{\mu^{eff}}{\mu^{eff}_{c}}, \quad \upsilon(\xi)=m(r)A^{3}/4\pi\mu^{eff}_{c},
 \end{split}
  \end{equation}
the subscript $c$ describes that the corresponding quantities are calculated at the center of the sphere. We obtain two equations Eqs.\eqref{61a} and \eqref{62a} of first order with three unknown   functions $\Psi$, $\upsilon$ and $\Pi$ which depend on two parameters $n$ and $\alpha$. One extra condition is required to model the system, which is established from vanishing complexity factor condition as
 \begin{eqnarray}\nonumber
   \frac{d\Pi}{d\xi}+\frac{3\Pi}{\xi} = \frac{1}{(8\pi+\frac{f_{T}}{2})}\left[4\pi\mu^{eff}_{c}n\Psi^{n-1}\frac{d\Psi}{d\xi}\right.\\\label{64a}
   \left.-\frac{3}{\xi}\left(L^{(D)}+4\pi^{(D)}\right)-L_{, \xi}^{(D)}-4\pi \frac{d\Pi^{(D)}}{d\xi}-\frac{\Pi f_{T,\xi}}{2}\right].
 \end{eqnarray}
 It gives us three ordinary differential Eqs.\eqref{61a}, \eqref{62a} and \eqref{64a} with three unknown functions $\Psi$, $\upsilon$ and $\Pi$.

 From Newtonian to general relativistic case we have two cases for polytropic equation of state, one of them is discussed in Eq.\eqref{60a} and the other is given below
 \begin{equation}\label{67a}
 P^{eff}_{r}=K\mu^{eff(\gamma)}_{b}=K\mu^{eff(1+1/n)}_{b},
 \end{equation}
 where $\mu^{(eff)}_{b}$ indicates the baryonic mass density. For this case, the equations identical to Eqs. \eqref{61a} and \eqref{64a} are
\begin{eqnarray}\nonumber
\xi^{2}\frac{d\Psi_{b}}{d\xi}\left[\frac{1-2\upsilon \alpha (n+1)/\xi}{1+\alpha\Psi_{b}}\right]+\upsilon+\alpha\xi^{3}\Psi^{n+1}_{b}
+\frac{2\Pi\xi\Psi^{-n}_{b}}{P^{eff}_{rc}(n+1)}~\\\label{68a}
~~~~\times\left[\frac{1-2\upsilon \alpha (n+1)/\xi}{1+\alpha\Psi_{b}}\right]=e^{\lambda}Z\left[\frac{\xi^{2}
(1-2\upsilon \alpha (n+1)/\xi)}{P^{eff}_{rc}A\Psi^{n}_{b}(1+\alpha\Psi_{b})(1+n)}\right],\\\nonumber
\frac{d\Pi}{d\xi}+\frac{3\Pi}{\xi}=\frac{1}{(8\pi+\frac{f_{T}}{2})}\left[4\pi\left(\mu^{(eff)}_{bc}n\Psi^{n-1}_{b}\frac{d\Psi_{b}}{d\xi}\left(1+K(n+1)
\mu^{(eff)\frac{1}{n}}_{bc}\Psi_{b}\right)\right)\right.\\\label{69a}-\left.\frac{3}{\xi}\left(L^{(D)}+4\pi^{(D)}\right)-L_{, \xi}^{(D)}-4\pi
 \frac{d\Pi^{(D)}}{d\xi}-\frac{\Pi f_{T,\xi}}{2}\right],
\end{eqnarray}
with $\Psi^{n}_{b}=\frac{\mu^{eff}_{b}}{\mu^{eff}_{bc}}$. Equations \eqref{68a} and \eqref{69a} explain the TOV equation and vanishing complexity condition with dimensionless variables in the $f(G,T)$ gravity,  respectively in the second case. The above equation describes the structure of spherical relativistic system with extra degrees of freedom mediated by $f(G,T)$ with different specific era of the cosmos controlled by an EoS parameter $K$. For instance,
\begin{enumerate}
  \item The choice $K=0$ in Eq.\eqref{68a} describes TOV equation for non-relativistic matter configurations.
  The condition $K=0$ in Eq.\eqref{60a} corresponds that $p^{eff}_{r}=0$, which means that dark
  source terms vanish out and we obtain a pressureless matter.
  Its so happened that one can analyze the hydrostatic equilibrium of a spherical object with the
  help of TOV equation. Furthermore, the input of $K=0$ in Eq.\eqref{69a} describes the equation of
  non-relativistic cloud having zero measure of complexity in it. However, the simultaneous solutions of Eqs.\eqref{68a} and \eqref{69a} could provide fruitful information for those non-relativistic spherical system that evolves non-adiabatically under the $f(G,T)$ gravity.
  \item One can analyze the role of dark energy as well $f(G,T)$ corrections on the existence of the homogeneous relativistic spheres with the help of Eqs.\eqref{68a} and \eqref{69a} after considering $K=-1$. Thus one can investigate the structural evolution of the regular distribution of matter content with the effects of cosmic inflation and accelerated expansion from the above two equations.
  \item Its could be possible that the spherical system could enter into the ultra-relativistic radiation phase. Thus, such a condition of the static irrotational spherical structure can be analyzed by keeping $K=\frac{1}{3}$ in Eqs.\eqref{68a} and \eqref{69a}.
      \item While the selection of $K\neq-1$ describes the set of equations of motion for a relativistic stellar structure in the phantom era of the universe with $f(G,T)$ corrections.
\end{enumerate}
Another interesting feature of the above equations is that one can properties of analyze various regular objects at different phases of the universe by keeping specific values to the polytropic index $n$. For example, the choice $n=1.5$ explains the polytropes for fully convective star cores \cite{chandrasekhar1957introduction, hansen2012stellar}, and $n=3$ demonstrates the core of massive white dwarfs \cite{sagert2006compact} and $n=\infty$ corresponds to isothermal sphere \cite{henrich1941stellar}.

 \section*{V. Conclusion}

This paper is devoted to comprehend the consequences of $f(G,T)$ gravity on the dynamics of self-gravitating spherical objects. Self
gravitating fluids in the field of astrophysics have such compelling
characteristics which create curiosity among researchers to analyze
their physical properties, like pressure, density, temperature,
stability etc. For this reason, we have analyzed static spherical
symmetric-metric associated with anisotropic matter. In the
formalism of $f(G,T)$ gravity, modified field and hydrostatic
equilibrium equations are formulated. Specific relations between $m$
and $m_{T}$ are developed by using the formalism given by
Misner-Sharp and Tolman, respectively. We have also performed
complete orthogonal splitting of Riemann curvature tensor in
$f(G,T)$ theory. Such a splitting lead us to get set of modified
versions of scalar variables that wouls have a direct relations with
the basic structural properties of the relativistic irrotational
spherical object. It is seen here that like GR, the orthogonal
breaking of Riemann tensor give rise to five scalar functions
$X_{TF},~X_{T},~Y_{TF},~Y_{T}$ and $Z_{T}$ in $f(G,T)$ gravity.
Herrera \cite{herrera2018new} introduced the idea of complexity for
self-gravitating anisotropic fluids. The elementary supposition is
that the system with isotropic pressure and homogenous energy
density is less complex. Among the derived scalars variables,
$Y_{TF}$ has been identified to be the complexity factor.
We found some results that can be expressed as under.\\

(i) The scalar $Y_{TF}$ covers the energy density inhomogeneity and pressure anisotropy under the effect of extra
curvature terms of $f(G,T)$ gravity, in a defined way.\\

(ii) Tolman mass in the presence of anisotropic pressure and inhomogeneous fluid is expressed in term
of this scalar along with dark source term under $f(G,T)$ gravity.\\

(iii) For the non-static dissipative case, $Y_{TF}$ contributes the dissipative fluxes in addition to
inhomogeneous energy density and local anisotropic pressure in the presence of modified terms.\\

(iv) The key reason for such a hypothesis lies in the fact that the scalar function $Y _{TF}$
includes influences from the inhomogeneity of energy density and the combined local pressure
anisotropy in a very unique way, which disappears completely for the regular and locally
isotropic distribution of fluids. It is important to note that the so-called complexity
factor not only dissolves for the aforementioned simple configuration, but can also disappear
when the two terms appearing in its definition, and featuring inhomogeneity of density and
anisotropic pressure, nullify each other. Thus complexity can lead to some various varieties
of systems, as mentioned in \cite{lopez1995statistical}.\\

After developing modified field equations and matter distribution,
we calculated the complexity of the system through one of the scalar
variables $Y_{TF}$ given in Eq.\eqref{43a} taken as complexity
factor which includes the contribution of energy density
inhomogeneity, anisotropic pressure and extra curvature terms.
Moreover, we present two applications of field equations by taking
the supposition of disappearing complexity factor condition i.e.
$Y_{TF}=0$. In the first example, we have examined the effects of
dark source terms by assuming specific energy density described by
Gookhro and Mehra ans\"{a}tz. In the second case, we have discussed
the polytropic equation of state and adopted new dimensionless
variables in order to write TOV equation, mass function and
complexity factor condition. This work would enable us to understand
the impact of complexity factor for self-gravitating objects in this
modified theory.  All the results of this modified theories are
minimized to GR by using the limit $f(G,T)=R$ \cite{herrera2018new}.

\vspace{0.5cm}

{\bf Acknowledgments}

\vspace{0.25cm}

The work of ZY and MZB were supported by National Research Project
for Universities (NRPU), Higher Education Commission, Islamabad
under the research project No. 8754. The work of KB was supported in part by the JSPS KAKENHI Grant Number JP
25800136 and Competitive Research Funds for Fukushima University Faculty
(19RI017).

  \end{document}